\documentstyle[aps,preprint]{revtex}
\def\lsim{\mathrel{\raise2pt\hbox to 8pt{\raise -5pt\hbox{$\sim$}\hss{$<$}}}}

\input BoxedEPS
\SetepsfEPSFSpecial
\begin{document}

\title{Implications of Pseudospin Symmetry on Relativistic Magnetic Properties and Gamow - Teller
Transitions in Nuclei}
\author{ {Joseph N. Ginocchio}} \address {Theoretical Division, Los Alamos National Laboratory, Los Alamos, New
Mexico 87545, USA}
\maketitle

\begin{abstract}
Recently it has been shown that pseudospin symmetry has its origins in a relativistic symmetry of the Dirac Hamiltonian.  Using this symmetry we relate
single - nucleon relativistic magnetic moments of states in a pseudospin doublet to the relativistic magnetic dipole transitions between the states in
the doublet, and we relate single - nucleon relativistic Gamow - Teller transitions within states in the
doublet. We apply these relationships to the Gamow - Teller transitions from $^{39}Ca$ to its mirror
nucleus $^{39}K$. 
\end{abstract}

\pacs{21.10.-k, 24.10.Jv, 21.60.Cs, 21.30.Fe}

\maketitle

\section{Introduction}

For nucleons moving in a relativistic mean field with scalar $V_S$ and vector potentials $V_V$, an SU(2) symmetry exists for the case
for which $V_S = - V_V$ \cite {bell}. This symmetry manifests itself in nuclei as a slightly broken
symmetry \cite {gino,gino2,ami,gino3} since $|{V_S + V_V
\over V_S - V_V}|$ is small for realistic mean fields \cite {wal,mad,cohen,arima,ring}, and, in fact, gives rise
to what has been called ``pseudospin symmetry''.  The original observations that led to the
coining of the word ``pseudospin symmetry" were quasi-degeneracies in spherical shell model orbitals with
non - relativistic quantum numbers ($n_r$,
$\ell$, $j =
\ell + 1/2)$ and ($n_{r}-1, \ell + 2$, $j = \ell + 3/2$) where $n_r$, $\ell$, and $j$ are 
the single-nucleon radial, orbital, and total 
angular momentum quantum numbers, respectively
\cite {kth,aa}.  This doublet structure is expressed in
terms of a ``pseudo'' orbital angular momentum 
$\tilde{\ell}$ = $\ell$ + 1, the average of the orbital angular momentum of the two 
states in doublet, and ``pseudo'' spin, $\tilde s$ = 1/2.
For example,
$(n_r s_{1/2},(n_r-1) d_{3/2})$ will have
$\tilde{\ell}= 1$ , $(n_r p_{3/2},(n_r-1) f_{5/2})$ will have $\tilde{\ell}= 2$, etc. 
These doublets are almost degenerate with
respect to pseudospin, since $j = \tilde{\ell}\ \pm \tilde s$ for the two states 
in the doublet; examples are shown in Figure 1. 
Pseudospin ``symmetry'' was shown
to exist in deformed nuclei as well \cite {bohr,draayer3} and has been used to explain
features of deformed nuclei, including superdeformation \cite {dudek} and identical bands
\cite {twin,stephens}. However, the origin of pseudospin symmetry remained a mystery and ``no deeper understanding of the origin of
these (approximate) degeneracies'' existed
\cite {ben}. A few years ago it was shown that relativistic mean field theories gave 
approximately the correct spin orbit splitting to
produce the pseudospin doublets \cite {draayer}. Finally the source of pseudospin symmetry as a broken symmetry of the Dirac Hamiltonian 
related to $V_S \approx - V_V$ was pointed out \cite {gino,gino2,ami,gino3}. For spherical nuclei,
pseudo-orbital angular momentum $\tilde{\ell}$ is also conserved and physically is the ``orbital
angular momentum'' of the lower component of the Dirac wavefunction.

One consequence of this relativistic SU(2) pseudospin symmetry is that the
spatial wavefunction for the lower component of the Dirac wavefunctions will be equal in shape
and magnitude for the two states in the doublet \cite {gino2,ami,gino3}.  For
spherical nuclei, this means that the radial wavefunctions for the lower components in the doublet will have the same number of
nodes, so we label these states with pseudo-radial quantum number (i.e.; the radial quantum number of the
lower component ($\tilde {n} = 0, 1,\dots$)).  Furthermore,the pseudo-orbital angular momentum will be a
conserved quantum number for spherical symmetric scalar and vector potentials and so we label the states
with the pseudo-orbital angular momentum 
$\tilde{\ell}$ \cite {ami}. Finally, the total angular momentum $ j\  (\vec{j} = \vec{\tilde{\ell}} + \vec {\tilde{1/2}}),$ 
and projection $m$, are
conserved as well. The Dirac wavefunction for the two states in the doublet are 
$$
\Psi_{\tilde {n},\tilde {\ell}, j = \tilde {\ell} + 1/2 , m} =
(g_{\tilde {n}-1, \tilde {\ell}, j } [Y_{\tilde {\ell} + 1}\chi]_m^{j = \tilde {\ell} + 1/2 },
if_{\tilde {n},\tilde {\ell},j} [Y_{\tilde {\ell}}\chi]_m^{j = \tilde {\ell} + 1/2 }),$$
\begin{equation}
\Psi_{\tilde {n}, \tilde {\ell}, j = \tilde {\ell} - 1/2, m} = (g_{\tilde {n}, \tilde {\ell}, j} [Y_{\tilde {\ell} - 1}
\chi]_m^{(j = \tilde {\ell} - 1/2)}, if_{\tilde {n}, \tilde {\ell}, j}[Y_{{\tilde
{\ell}}}\chi]_m^{(j = \tilde {\ell} - 1/2)}), 
\label {dirac}
\end{equation}
where $g, f$ are the radial wave functions, 
$Y_{\tilde {\ell}}$ are the spherical harmonics,  $\chi$ is a two-component Pauli spinor, and
$[\dots]^{(j)}$ means coupled to angular momentum $j$. We note that the upper component of the $j= \tilde {\ell} - 1/2 $ wavefunction has the same
radial quantum number as the lower component, whereas the upper component of the $j= \tilde {\ell} + 1/2 $ wavefunction has 
radial quantum number one unit less than the lower component. The normalization of the wavefunction gives
$$
\int_0^{\infty} [ g_{\tilde {n}^{\prime}, \tilde {\ell}, j}^2  + f_{\tilde {n}, \tilde {\ell}, j}^2] r^2 dr = 1;
$$
\begin{equation}
j = \tilde {\ell} + 1/2, \ \tilde {n}^{\prime} = \tilde {n} - 1; \ j = \tilde {\ell} - 1/2, \ \tilde {n}^{\prime} = \tilde {n}.
\label{norm}
\end{equation}

For a square well potential, the overall phase between the two amplitudes
will be a minus sign
\cite {gino} so we expect that, in the symmetry limit for realistic potentials, $f_{\tilde {n}, \tilde {\ell}, j = \tilde {\ell} + 1/2}(r) = -
f_{\tilde {n},
\tilde {\ell}, j = \tilde {\ell} - 1/2}(r) = f_{\tilde {n}, \tilde {\ell}}(r) $.  For the relativistic mean field
approximation to relativistic Lagrangrians with realistic zero range interactions and to nuclear field theory with meson exchanges it was indeed shown
that,
$f_{\tilde {n},
\tilde {\ell}, j = \tilde {\ell} + 1/2}(r) \approx -  f_{\tilde {n}, \tilde {\ell}, j = \tilde
{\ell} - 1/2}(r)$
\cite {gino2,ring}.  

However, to date, the effect of pseudospin symmetry on the relativistic wavefunction has not been tested empirically. Since the lower component
of the Dirac wavefunction is small \cite {gino2,gino3,ring} this effect will be difficult to detect except perhaps in certain forbidden transitions.
For example, single - nucleon magnetic dipole and Gamow-Teller transitions between pseudospin doublets are forbidden non-relativistically (i.e.,
``$\ell$ forbidden" \cite {ian2,achim}) because the orbital angular momenta of the two states differ by two units. However, they are not forbidden
relativistically. In this paper we shall use  approximate pseudospin symmetry in the wavefunction to derive relations between single-nucleon
relativistic magnetic moments and magnetic dipole transtions within a pseudospin doublet on the one hand, and between single-nucleon relativistic
Gamow-Teller transitions within a pseudospin doublet on the other hand. These relationships provide a test for the influence of pseudospin
symmetry on the single - nucleon wavefunctions.

\section{Magnetic Moments and Transitions}

The relativistic magnetic dipole operator for a particle with charge e is given by \cite {henry,miller},

\begin{equation}
\hat {\mu}_i =  -{e\over 2}\ g_{\rho}\ ({{\vec{\alpha}} \times{\vec r}})_i + \mu_{A,\rho}\ 
\sigma_i,
\label {mm}
\end{equation}
where ${\vec \alpha}$ is the usual Dirac matrix, 
$\vec r$ is the three space vector, $\rho = \pi$ for a proton and $\nu$ for a neutron, $g_{\rho}$ is the orbital gyromagnetic ratio,
$g_{\pi} = 1, g_{\nu} = 0$, and $\mu_{A,\rho}$ is the anamolous magnetic moment, $\mu_{A,\pi} = 1.793\mu_0,
\ \mu_{A,\nu} = -1.913\mu_0$, where $ \mu_0 = {e\hbar \over 2Mc}$ is the nuclear magneton. The magnetic moment is
given in terms of the matrix element of this operator with m = j,
\begin{equation}
 {\mu}_{j,\rho} =  
\langle \Psi_{{\tilde {n}},{\tilde {\ell}}, { {j}},m = j, \rho} | \hat {\mu} | \Psi_{{\tilde {n}}, \tilde {\ell},j, m = j,\rho }\rangle,
\label {mm}
\end{equation}
and the square root of the magnetic transition probability between two states in the doublet is given in terms of the reduced matrix element of
this operator,
\begin{equation}
\sqrt{B(M1:{\tilde {n}},{\tilde {\ell}},j^{\prime} \rightarrow {\tilde {n}},{\tilde {\ell}},j )_{\rho}} = {1\over (2 j^{\prime} + 1)}\langle
\Psi_{{\tilde {n}}^{\prime},{\tilde {\ell}}, { {j}^{\prime}}, \rho} ||\hat {\mu} ||\Psi_{{\tilde {n}},
\tilde {\ell},j, \rho }\rangle
\label {bm}
\end{equation}
Using the Dirac wavefunction (\ref {dirac}), this results in
$$
\ j = \tilde {\ell} - 1/2
$$
\begin{equation}
\mu_{j, \rho} = \\ {- e\ g_{\rho}\ (j + 1/2)\over 2(j+1)}  \int_0^{\infty} g_{\tilde {n}, \tilde {\ell}, j
}f_{\tilde {n},\tilde {\ell},j,\rho} \ r^3\ dr + \mu_{A,\rho}\ ( 1 - {(2j+1)\over (j+ 1)}  \ \int_0^{\infty} f_{\tilde {n},
\tilde {\ell}, j,\rho }^2 \ r^2\ dr)\ ,
\label {mm1}
\end{equation}
$$
\ j = \tilde {\ell} + 1/2
$$
\begin{equation}
\mu_{j, \rho} = \\ {e\ g_{\rho} (j + 1/2)\over 2(j+1)}  \int_0^{\infty} g_{\tilde {n}-1, \tilde {\ell}, j, \rho
}f_{\tilde {n},\tilde {\ell},j, \rho} \ r^3\ dr - {\mu_{A,\rho}\over (j+ 1)}\ ( j - (2j+1)  \ \int_0^{\infty} f_{\tilde {n},
\tilde {\ell}, j, \rho }^2 \ r^2\ dr)\  ,
\label {mm2}
\end{equation}
$$
\ {j}^{\prime} = \tilde {\ell} + 1/2, \  j = \tilde {\ell} - 1/2
$$
$$ 
\sqrt{B(M1:{\tilde {n}},{\tilde {\ell}}, j^{\prime} \rightarrow {\tilde {n}},{\tilde {\ell}}, j )_{\rho}} =  -\sqrt{{(2j + 1) }\over {( 2j + 3)
}}\sqrt{B(M1:{\tilde {n}},{\tilde {\ell}}, j \rightarrow {\tilde {n}},{\tilde {\ell}}, j^{\prime} )_{\rho}} = 
$$
\begin{equation}
-{1\over 4} \sqrt{{(2j + 1)
}\over {( j + 1) }}\ [ {{e \ g_{\rho}\over 2 }} \ 
\int_0^{\infty} [g_{\tilde {n}-1,
\tilde {\ell}, j^{\prime},\rho }f_{\tilde {n},\tilde {\ell},j,\rho} +  g_{\tilde {n}, \tilde
{\ell}, j, \rho }f_{\tilde {n},\tilde {\ell},j^{\prime}, \rho}] \ r^3\ dr 
+ 4 \mu_{A,\rho}\ \int_0^{\infty} f_{\tilde {n}, \tilde
{\ell}, j^{\prime}, \rho }f_{\tilde {n},\tilde {\ell},j,\rho}\ r^2\ dr\ ].
\label {bm1}
\end{equation}

\subsection{Non-relativistic Limit}

The Dirac equation with speherically symmetric potentials reduces to two coupled one - dimensional radial equations for the upper and lower
components, $(g,f)$ \cite {gino},
\begin{equation}
\hbar \ c[ {d\over dr} + {{1 + \kappa}\over r} ] g_{\tilde {n}^{\prime}, \tilde {\ell}, j,
\rho} =  [2Mc^2 - E + V_S - V_V]\ f_{\tilde {n}, \tilde {\ell}, j, \rho},
\label {diraceqn1}
\end{equation}
\begin{equation}
\hbar\ c [ {d\over dr} + {{1 - \kappa}\over r} ] f_{\tilde {n}, \tilde {\ell}, j,
\rho} = [ E + V_S + V_V]\ g_{\tilde {n}^{\prime}, \tilde {\ell}, j, \rho},
\label {diraceqn2}
\end{equation}
where 
\begin {equation}
\kappa = - \tilde {\ell}  , j = \tilde {\ell} - 1/2;\ \  \kappa = \tilde {\ell} + 1 , j = \tilde
{\ell} + 1/2,
\label {kappa}
\end{equation}  
$M$ is the nucleon mass, and $E$ is the binding energy. 
In order to
determine 
$\int_0^{\infty} gf r^3\ dr$ we use (\ref {diraceqn1}, \ref {diraceqn2}) to derive \cite {henry}:
\pagebreak
$$
g_{\tilde {n}^{\prime}, \tilde {\ell}, j^{\prime},\rho}f_{\tilde {n}, \tilde {\ell}, j,
\rho} = 
$$
\begin{equation}
{\hbar c \over {2Mc^2 + 2V_S}} \ [g_{\tilde {n}^{\prime}, \tilde {\ell},
j^{\prime},\rho}{d\over dr}g_{\tilde {n}^{\prime}, \tilde {\ell}, j,\rho} + f_{\tilde {n},
\tilde {\ell}, j^{\prime},\rho}{d\over dr}f_{\tilde {n}, \tilde {\ell}, j,\rho}+ {{1
+ \kappa}\over r} g_{\tilde {n}^{\prime}, \tilde {\ell}, j^{\prime},\rho} g_{\tilde {n}^{\prime}, \tilde
{\ell},j,\rho} +  {{1 -
\kappa}\over r} f_{\tilde {n}, \tilde {\ell}, j^{\prime},\rho}f_{\tilde {n}, \tilde
{\ell}, j,\rho}]
\label {fg}
\end{equation}
In the non-relativistic limit, the potentials are ignored with respect to the nucleon
mass, although ${V_S \over Mc^2 }\approx .48$ in the interior of the nucleus. 
Also terms quadratic in f are ignored. This gives
\begin{equation}
\int_0^{\infty} r^3\ dr [g_{\tilde {n}^{\prime}, \tilde {\ell}, j^{\prime},\rho}f_{\tilde
{n}, \tilde {\ell}, j,\rho} + g_{\tilde {n}^{\prime}, \tilde {\ell}, j,\rho}f_{\tilde
{n}, \tilde {\ell}, j^{\prime},\rho} ] = {\hbar \over 2Mc}(\kappa + {\kappa}^{\prime}
- 1) \ \int_0^{\infty} r^2\ dr g_{\tilde {n}^{\prime}, \tilde {\ell}, j^{\prime},\rho}\
g_{\tilde {n}^{\prime}, \tilde {\ell}, j,\rho}.
\end{equation}
For $j^{\prime} = j,\ \int_0^{\infty} r^3\ dr g_{\tilde {n}^{\prime}, \tilde {\ell}, j,\rho}\
g_{\tilde {n}^{\prime}, \tilde {\ell}, j,\rho}$ =1 from the normalization condition (\ref {norm}). 
Therefore in the non-relativistic limit, the magnetic moments become,
\begin{equation}
\mu_{j, \rho} = \\  (j +1/2)\ g_{\rho}\ \mu_0 + \mu_{A,\rho}; \ j = \tilde {\ell} -
1/2,
\label{nr1}
\end{equation}

\begin{equation}
\mu_{j, \rho} = \\ { j \over (j+1)} ( (j + 1/2)\ g_{\rho}\ \mu_0  -
\mu_{A,\rho}); \ j = \tilde {\ell} + 1/2.
\label {nr2}
\end{equation}
The non-relativistic limits
for the magnetic moments in (\ref {nr1},\ref {nr2}) are equivalent to the Schmidt values
\cite {herman}.

However, for $j^{\prime}\ne
j$, it follows from (\ref {kappa}) that $\kappa + {\kappa}^{\prime} -1 = 0$ and therefore,
\begin{equation}
 {B(M1:{\tilde {n}},{\tilde {\ell}}, j^{\prime} \rightarrow {\tilde {n}},{\tilde
{\ell}}, j )_{\rho} = 0; \  {j}^{\prime}} \ne j,
\label {nr3}
\end{equation}
Thus the non-relativistic limit of the B(M1) is zero which is as it should
be since the transition is from $\ell$ to $\ell \pm 2 $  as stated in the
Introduction.

\subsection{Pseudospin Symmetry }

Instead of looking at the non-relativistic limit, we examine the pseudospin limit
which assumes that the spatial wave functions of the lower components of the doublet
are equal and opposite in sign,
\begin{equation}
f_{\tilde {n}, \tilde {\ell}, j = \tilde {\ell} + 1/2,\rho} (r) = - f_{\tilde {n},
\tilde {\ell}, j = \tilde {\ell} - 1/2, \rho}(r) = f_{\tilde {n}, \tilde {\ell},\rho}(r). 
\label {ps}
\end{equation}
Inserting this relation into (\ref {mm1}, \ref{mm2}, \ref{bm1}) we obtain,
$$
\ j = \tilde {\ell} - 1/2
$$
\begin{equation}
\mu_{j, \rho} = \\ { e\ g_{\rho}\ (j + 1/2)\over 2(j+1)}  \int_0^{\infty} g_{\tilde {n},
\tilde {\ell}, j, \rho }f_{\tilde {n},\tilde {\ell},\rho} \ r^3\ dr + \mu_{A,\rho}\ ( 1 -
{(2j+1)\over (j+ 1)}  \
\int_0^{\infty} f_{\tilde {n},
\tilde {\ell}, \rho }^2 \ r^2\ dr)\ ,
\label {psmm1}
\end{equation}
$$
\ j = \tilde {\ell} + 1/2
$$
\begin{equation}
\mu_{j, \rho} = \\ {e\ g_{\rho} (j + 1/2)\over 2(j+1)}  \int_0^{\infty} g_{\tilde {n} - 1, \tilde {\ell}, j, \rho
}f_{\tilde {n},\tilde {\ell},\rho} \ r^3\ dr - {\mu_{A,\rho}\over (j+ 1)}\ ( j - (2j+1)  \
\int_0^{\infty} f_{\tilde {n},
\tilde {\ell}, \rho }^2 \ r^2\ dr)\  ,
\label {psmm2}
\end{equation}
$$
\ {j}^{\prime} = \tilde {\ell} + 1/2, \  j = \tilde {\ell} -
1/2
$$
$$ 
\sqrt{B(M1:{\tilde {n}},{\tilde {\ell}}, j^{\prime} \rightarrow {\tilde {n}},{\tilde {\ell}}, j )_{\rho}} =  -\sqrt{{(2j + 1) }\over {( 2j + 3)
}}\sqrt{B(M1:{\tilde {n}},{\tilde {\ell}}, j \rightarrow {\tilde {n}},{\tilde {\ell}}, j^{\prime} )_{\rho}} = 
$$
\begin{equation}
-{1\over 4} \sqrt{{(2j + 1)
}\over {( j + 1) }}\ [ {{e \ g_{\rho} }\over 2} \ 
\int_0^{\infty} [- g_{\tilde {n}-1,
\tilde {\ell}, j^{\prime},\rho } +  g_{\tilde {n}, \tilde
{\ell}, j, \rho }]f_{\tilde {n},\tilde {\ell},\rho}] \ r^3\ dr 
- 4 \mu_{A,\rho}\ \int_0^{\infty} f_{\tilde {n}, \tilde
{\ell},  \rho }^2\ r^2\ dr\ ].
\label {psbm1}
\end{equation}

For neutrons $g_{\nu} = 0$, and hence we have one unkown quantity, $\int_0^{\infty} f_{\tilde {n}, \tilde
{\ell},  \rho }^2\ r^2\ dr$.  Therefore, if we know one magnetic quantity,
we can predict two others,
\begin{equation}
 \sqrt{B(M1:{\tilde {n}},{\tilde {\ell}}, j^{\prime} \rightarrow {\tilde
{n}},{\tilde {\ell}}, j} )_{\nu} = - \sqrt{j + 1\over 2j + 1}( \mu_{j,\nu} -
\mu_{A,\nu}),
\label {neutron} 
\end{equation}
\begin{equation}
 \sqrt{B(M1:{\tilde {n}},{\tilde {\ell}}, j^{\prime} \rightarrow {\tilde
{n}},{\tilde {\ell}}, j} )_{\nu} =  {j + 2 \over 2j + 3}\  \sqrt{2j +
1\over j + 1}(\mu_{j^{\prime},\nu} + {{j+1}\over {j + 2}} \mu_{A,\nu}).
\label {neutron2} 
\end{equation}

For protons there are three unkown integrals, and so we can only derive one
relationship between the three magnetic quantities,
$$
\sqrt{B(M1:{\tilde {n}},{\tilde {\ell}}, j^{\prime} \rightarrow {\tilde
{n}},{\tilde {\ell}}, j} )_{\pi} =
 {((j + 2) (2j + 1)\mu_{j^{\prime},\pi} -
(2j + 3) (j + 1)\mu_{j,\pi} + 4\ (j + 1)^2\ \mu_{A,\pi})\over \ 2\ (2j +
3)\sqrt{(j + 1) (2j + 1)}};
$$
\begin{equation}
\ j^{\prime} = \tilde {\ell} +
1/2, \ j = \tilde {\ell} - 1/2.
\label {proton} 
\end{equation}

If the magnetic moments are given by the Schmidt values as in (\ref {nr1}, \ref{nr2}), then the magnetic
transitions in (\ref {neutron}, \ref {neutron2}, \ref {proton}) will be identically zero, which is consistent with
the non-relativistic limit.  

The relativistic mean field overestimates the isoscalar magnetic moments of nuclei \cite
{miller}.  However, when the response of the spectator nucleons is included, the relativistic isoscalar magnetic
moments agree better with experiment \cite {ralph}. The response of the spectator nucleons do not significantly
affect isovector magnetic moments since the dominant mesons in the relativistic field theory are isoscalar.  If we
define the isoscalar and vector operators as 
$$
\mu_{j,S} = {1\over 2}(\mu_{j,\nu} + \mu_{j,\pi}); \mu_{j,V} = {1\over 2}(\mu_{j,\nu} - \mu_{j,\pi}); \mu_{A,S} =
{1\over 2}(\mu_{A,\nu} + \mu_{A,\pi}); \mu_{A,V} = {1\over 2}(\mu_{A,\nu} - \mu_{A,\pi}); 
$$
$$
\sqrt{B(M1:{\tilde {n}},{\tilde {\ell}}, j^{\prime} \rightarrow {\tilde
{n}},{\tilde {\ell}}, j} )_{S} = {1\over 2} (\sqrt{B(M1:{\tilde {n}},{\tilde {\ell}}, j^{\prime} \rightarrow {\tilde
{n}},{\tilde {\ell}}, j} )_{\nu} + \sqrt{B(M1:{\tilde {n}},{\tilde {\ell}}, j^{\prime} \rightarrow {\tilde
{n}},{\tilde {\ell}}, j} )_{\pi});
$$
\begin{equation}
\sqrt{B(M1:{\tilde {n}},{\tilde {\ell}}, j^{\prime} \rightarrow {\tilde
{n}},{\tilde {\ell}}, j} )_{V} = {1\over 2} (\sqrt{B(M1:{\tilde {n}},{\tilde {\ell}}, j^{\prime} \rightarrow {\tilde
{n}},{\tilde {\ell}}, j} )_{\nu} - \sqrt{B(M1:{\tilde {n}},{\tilde {\ell}}, j^{\prime} \rightarrow {\tilde
{n}},{\tilde {\ell}}, j} )_{\pi}),
\label {SV} 
\end{equation}
then the relations are separated into relations among the isoscalar and isovector magnetic properties:
$$
\sqrt{B(M1:{\tilde {n}},{\tilde {\ell}}, j^{\prime} \rightarrow {\tilde
{n}},{\tilde {\ell}}, j} )_{S/V} =
 {((j + 2) (2j + 1)\mu_{j^{\prime},S/V} -
(2j + 3) (j + 1)\mu_{j,S/V} + 4\ (j + 1)^2\ \mu_{A,S/V})\over \ 2\ (2j +
3)\sqrt{(j + 1) (2j + 1)}};
$$
\begin{equation}
\ j^{\prime} = \tilde {\ell} +
1/2, \ j = \tilde {\ell} - 1/2.
\label {S/Vr} 
\end{equation}

\section{Gamow - Teller Tansitions}

The Gamow - Teller operator is given by
\begin{equation}
GT = {g_A \over \sqrt{2}}\ \sigma \tau_{\pm},
\label {GT} 
\end{equation}
where $g_A$ is the axial vector coupling constant (= 1.2670 (35)) and $\tau_{\pm}$ are the isospin raising and lowering
operator. Thus this operator is a pure isovector operator.
Using the Dirac wavefunction (\ref {dirac}), this results in
$$
\ j = \tilde {\ell} - 1/2
$$
\begin{equation}
\sqrt{B(GT:{\tilde {n}},{\tilde {\ell}}, j, \rho \rightarrow
{\tilde {n}},{\tilde {\ell}}, j, \bar{\rho} )} = \sqrt{(j + 1) \over j}\
g_{A}\ ( 1 - {(2j+1)\over (j+ 1)} 
\
\int_0^{\infty} f_{\tilde {n}, \tilde {\ell}, j,\rho }f_{\tilde {n}, \tilde
{\ell}, j,\bar{\rho} } \ r^2\ dr)\ ,
\label {gt1}
\end{equation}
$$
\ j = \tilde {\ell} + 1/2
$$
\begin{equation}
\sqrt{B(GT:{\tilde {n}},{\tilde {\ell}}, j, \rho \rightarrow
{\tilde {n}},{\tilde {\ell}}, j, \bar{\rho} )} = -{g_{A}\over{\sqrt{j (j+
1)}}}\ ( j - (2j+1)  \
\int_0^{\infty} f_{\tilde {n}, \tilde {\ell}, j, \rho } f_{\tilde {n},
\tilde {\ell}, j, \bar{\rho}} \ r^2\ dr)\  ,
\label {gt2}
\end{equation}
$$
\ {j}^{\prime} = \tilde {\ell} + 1/2, \  j = \tilde {\ell} - 1/2
$$
$$ 
\sqrt{B(GT:{\tilde {n}},{\tilde {\ell}}, j^{\prime}, {\rho} \rightarrow
{\tilde {n}},{\tilde {\ell}}, j, \bar{\rho} )} =  -\sqrt{{(2j + 1) }\over
{( 2j + 3) }}\sqrt{B(GT:{\tilde {n}},{\tilde {\ell}}, j,\bar{\rho} 
\rightarrow {\tilde {n}},{\tilde {\ell}}, j^{\prime}, {\rho} )} = 
$$
\begin{equation}
-\sqrt{{(2j + 1) }\over {j + 1 }}\   g_{A}\ \int_0^{\infty} f_{\tilde
{n}, \tilde {\ell}, j^{\prime}, \rho }f_{\tilde
{n},\tilde {\ell},j,\bar{\rho}}\ r^2\ dr\ .
\label {gt3}
\end{equation}
where $\bar{\rho} = \pi$ if $\rho = \nu$ and $\bar{\rho} = \nu$ if $\rho = \pi$.

We notice that 
\begin{equation}
\sqrt{B(GT:{\tilde {n}},{\tilde {\ell}}, j, \rho \rightarrow
{\tilde {n}},{\tilde {\ell}}, j, \bar{\rho} )} = \sqrt{B(GT:{\tilde {n}},{\tilde {\ell}},
j,\bar{\rho} \rightarrow {\tilde {n}},{\tilde {\ell}}, j, \rho )},
\label {gte}
\end{equation}
but, in general, 
\begin{equation}
\sqrt{B(GT:{\tilde {n}},{\tilde {\ell}}, j^{\prime}, \rho \rightarrow
{\tilde {n}},{\tilde {\ell}}, j, \bar{\rho} )} \neq \sqrt{B(GT:{\tilde {n}},{\tilde {\ell}},
j^{\prime},\bar{\rho} \rightarrow {\tilde {n}},{\tilde {\ell}}, j, \rho )},
\label {gtne}
\end{equation}
\subsection{Non-Relativistic Limit of the Gamow - Teller Transitions}
Since terms quadratic in $f$ are ignored in the non- relatvistic limit, we get the usual results,

\begin{equation}
\sqrt{B(GT:{\tilde {n}},{\tilde {\ell}}, j, \rho \rightarrow
{\tilde {n}},{\tilde {\ell}}, j, \bar{\rho} )} = \sqrt{(j + 1)\over j} \
g_{A}; \ j = \tilde {\ell} - 1/2,
\label {gtnr1}
\end{equation}

\begin{equation}
\sqrt{B(GT:{\tilde {n}},{\tilde {\ell}}, j, \rho \rightarrow
{\tilde {n}},{\tilde {\ell}}, j, \bar{\rho} )} = -\sqrt{j\over
(j + 1)}\ {g_{A}};\ j = \tilde {\ell} + 1/2
\label {gtnr2}
\end{equation}

\begin{equation} 
\sqrt{B(GT:{\tilde {n}},{\tilde {\ell}}, j^{\prime}, {\rho} \rightarrow
{\tilde {n}},{\tilde {\ell}}, j, \bar{\rho} )} = 0;\
{j}^{\prime}\ne j
\label {gtnr3}
\end{equation}

\subsection{Pseudospin Symmetry}

Using pseudospin symmetry, (\ref {ps}), there is only one unkown for the Gamow - Teller 
transtions and hence each transition is related to the other,
$$
\ j^{\prime} = \tilde {\ell} +
1/2, \ j = \tilde {\ell} - 1/2.
$$
\begin{equation}
 \sqrt{B(GT:{\tilde {n}},{\tilde {\ell}}, j^{\prime}, {\rho} \rightarrow {\tilde
{n}},{\tilde {\ell}}, j,\bar{\rho})}  = - \sqrt{j \over 2j + 1}( \sqrt{B(GT:{\tilde
{n}},{\tilde {\ell}}, j, {\rho} \rightarrow {\tilde {n}},{\tilde {\ell}}, j, \bar{\rho}
)}  - \sqrt{j + 1 \over j }\ g_A),
\label {gtps1} 
\end{equation}
\begin{equation}
 \sqrt{B(GT:{\tilde {n}},{\tilde {\ell}}, j^{\prime},{\rho} \rightarrow {\tilde
{n}},{\tilde {\ell}}, j,\bar{\rho})}  =  {\sqrt{(j + 2 ) (2j +
1)}  \over 2j + 3}\  (\sqrt{B(GT:{\tilde {n}},{\tilde {\ell}}, j^{\prime}, {\rho} \rightarrow {\tilde
{n}},{\tilde {\ell}}, j^{\prime}, \bar{\rho})}  + \sqrt{{{j+1}\over {j + 2}}} g_A),
\label {gtps2} 
\end{equation}
$$
\sqrt{B(GT:{\tilde {n}},{\tilde {\ell}}, j, {\rho} \rightarrow
{\tilde {n}},{\tilde {\ell}}, j,\bar{\rho}} ) = 
$$
\begin{equation}
- {(2j + 1) \over (2j +
3)}\ \sqrt {{j + 2} \over{j}} \ ( \sqrt{B(GT:{\tilde {n}},{\tilde {\ell}}, j^{\prime},{\rho} \rightarrow {\tilde
{n}},{\tilde {\ell}}, j^{\prime},\bar{\rho})} - {2\over (2j + 1)}\ \sqrt{{j + 1} \over {j + 2}}g_A), 
\label{gtps3}
\end{equation}
\begin{equation}
 \sqrt{B(GT:{\tilde {n}},{\tilde {\ell}}, j^{\prime},{\rho} \rightarrow {\tilde
{n}},{\tilde {\ell}}, j,\bar{\rho})}  =  \sqrt{B(GT:{\tilde {n}},{\tilde {\ell}}, j^{\prime}, \bar{\rho} \rightarrow
{\tilde {n}},{\tilde {\ell}}, j, {\rho})}. 
\label {gtps4} 
\end{equation}
This last relation, (\ref {gtps4}), also follows from isospin symmetry as well, but if pseudospin symmetry is conserved than the
relation holds even though isospin may be violated; i.e., $f_{\tilde {n}, \tilde {\ell}, \pi } \neq f_{\tilde {n}, \tilde
{\ell}, \nu }$.
\section{An example: $^{39}$K, $^{39}$Ca }
The nuclei $^{39}_{19}$K$_{20}$ and $^{39}_{20}$Ca$_{19}$ are mirror nuclei. The ground state and first excited state
of $^{39}_{19}$K$_{20}$ are interpreted as a $0d_{3/2}$ and $1s_{1/2}$ proton hole respectively, while the ground
state and first excited state of $^{39}_{20}$Ca$_{19}$ are interpreted as a $0d_{3/2}$ and $1s_{1/2}$ neutron hole
respectively. These states are members of the ${\tilde {n}} = 1, {\tilde {\ell}} =1$
pseudospin doublet. The M1 transitions between these two states in both of these nuclei
have been measured, although they are forbidden in a non-relativistic single-nucleon model, and are indeed small \cite {ian,ian2}.  The
magnetic moments of the ground states are known.  However, the magnetic moments of the excited states are not known
so the magnetic relationships introduced in (\ref {S/Vr}) can not be tested at this time.

On the other hand, the Gamow - Teller transitions from the ground state of $^{39}$Ca to the ground and first excited
state of $^{39}$K are known as indicated in Figure 2, which is enough information to test (\ref{gtps2}). For this example, $j =
1/2$, (\ref{gtps2}) beomes
\begin{equation}
 \sqrt{B(GT:{\tilde {1}},{\tilde {1}}, {3/2}^+,\nu \rightarrow {\tilde
{1}},{\tilde {1}}, {1/2}^+,\pi )} =  {\sqrt{5}  \over 4}\ (\sqrt{B(GT:{\tilde {1}},{\tilde {1}},
{3/2}^+, {\nu} \rightarrow {\tilde {1}},{\tilde {1}},{3/2}^+ , \pi )} + \sqrt{0.6}
g_A).
\label {Ca} 
\end{equation}

Of course only the $B(GT)$ is measured; the sign of the square root is unkown.  However, we choose the negative sign, 
$\sqrt{B(GT:{\tilde {1}},{\tilde {1}}, {3/2}^+, {\nu} \rightarrow {\tilde {1}},{\tilde {1}},{3/2}^+ , \pi} )_{exp} = - 0.647 (10) $ \cite
{ian}, because in the non-relativistic limit given in (\ref {gtnr2}), the square root is negative, which also agrees with shell
model calculations \cite {ian2}. Since we
are dealing with a single - nucleon model we can expect renormalization of the coupling constant $g_A$ due to omitted
shell model configurations just as in the non-relativistic shell model \cite {BW}. In
Table 1 we see that the quenching necessary to
reproduce the experimental ``$\ell$ forbidden'' transition $\sqrt{B(GT:{\tilde {1}},{\tilde {1}}, {3/2}^+,\nu \rightarrow {\tilde
{1}},{\tilde {1}}, {1/2}^+,\pi} )_{exp}$ is consistent with the quenching needed in the non-relativistic shell model to
reproduce $\ell$ allowed Gamow - Teller transitions. In the non-relativistic shell model an effective tensor term $g_{eff}\ [ Y_2 \sigma]^{(1)}$ is added
to the Gamow-Teller   operator, where $Y_2$ is the spherical harmonic of rank two and $[\dots]^{(1)}$ means coupled to angular momentum rank unity. Using
a calculated effective coupling constant $g_{eff}$ which includes core polarization, isobar excitations, meson exchange currents, and relativistic
corrections, a value of the  ``$\ell$ forbidden'' transition $\sqrt{B(GT:{\tilde {1}},{\tilde {1}}, {3/2}^+,\nu \rightarrow {\tilde
{1}},{\tilde {1}}, {1/2}^+,\pi} )_{NR} = -0.036(18)$ is calculated.  This value agrees with the experimental value within the limits of experimental and
theoretical uncertainty.  However, the isoscalar and isovector  magnetic dipole transitions calculated between the same states and using the same model
disagrees with the experimental transitions by a factor of four to five \cite {ian}. A measurement of the magnetic moments of the $s_{1/2}$ excited states
in $^{39}K$ and
$^{39}Ca$ would allow the prediction of the forbidden  magnetic dipole transitions via (\ref {S/Vr}) which may  be helpful in throwing light on this
dilemma. 

We can now predict the $1/2^+ \rightarrow 1/2^+$ transition using (\ref{gtps3}). The results are tabulated in Table 2; this
transition is the largest within the doublet. Furthermore, the final transition, which is also $``\ell$ forbidden'', can be
determined from (\ref {gt3}) and (\ref {gtps4}): 
$$
\sqrt{B(GT:{\tilde
{n}},{\tilde {\ell}}, j = 1/2^+, {\nu} \rightarrow {\tilde {n}},{\tilde {\ell}}, j^{\prime} = 3/2^+, \pi}) =
$$
\begin {equation}
-\sqrt{2}\sqrt{B(GT:{\tilde {n}},{\tilde {\ell}}, j = 3/2^+, {\nu} \rightarrow {\tilde {n}},{\tilde {\ell}},
j^{\prime} = 1/2^+,
\pi}) = \mp 0.034 (1).
\end{equation}
This relationship does not depend on the effective $g_A$ but also follows from isospin symmetry as well.

\section{Conclusions}

Recent investigations suggest that pseudospin symmetry appears to be only slightly broken particularly near the Fermi sea 
\cite {gino,gino2,ami,ring,arima,gino3}. The empirical evidence for pseudospin symmetry has been in the small energy splittings between
doublets. In this paper we analyzed magnetic dipole properties and Gamow-Teller transitions under assumption that pseudospin symmetry is conserved. 
Pseudospin conservation implies that the spatial wavefunctions of the lower component of the Dirac single - nucleon wavefunction are equal and
opposite in sign for pseudospin doublets.  Using this assumption, we derive, for spherical nuclei, a relationship for the scalar (vector)
magnetic dipole transition between the two states of the doublet and the scalar (vector) magnetic moments of the two states in the doublet.  Under the
same assumptions we derive relationships between any two Gamow-Teller transitions from states in the doublet to states in the doublet.  We applied the
Gamow-Teller relation to the ``$\ell$ forbidden''  $\beta$ - decay of $^{39}Ca$, and  conclude that agreement occurs for a quenching of the axial
coupling constant comparable to that neccessary to fit $\ell$ allowed Gamow-Teller transitions in the non-relativistic shell
model \cite {BW,B}. We point out that a measurement of the magnetic moments of the $s_{1/2}$ excited states in $^{39}K$ and $^{39}Ca$ would allow the
prediction of the forbidden  magnetic dipole transitions via (\ref {S/Vr}) which may  be helpful in throwing light on an inconsistency posed by the
non-relativistic shell model \cite {ian2}.  Furthermore we predict the other two Gamow-Teller transitions from the $1s_{1/2}, 1d_{3/2}$ states in
$^{39}Ca$ to their isobaric analogues in $^{39}K$ using pseudospin symmetry, thereby producing a test of the effect of pseudospin symmetry on the
relativistic single - nucleon wavefunctions.

\section{Acknowlegements}

The author would like to thank Anna Hayes and Ian Towner for discussions. This research is supported by the U. S. Department of Energy
under contract W-7405-ENG-36.\\

\begin{table}
\caption{Predicted ``$\ell$ forbidden'' Gamow - Teller strength, $^{39}Ca \rightarrow\  ^{39}K$, for various values of the
effective axial coupling constant.}
\vspace{18pt}
\vbox {\tabskip 2em plus 3em minus 1em \halign to \hsize{\hfil  
#\hfill && #\hfil \hfil 
\cr ${{\tilde g}_A} $ & $\sqrt{B(GT:{\tilde {1}},{\tilde {1}}, {3/2}^+,\nu \rightarrow {\tilde
{1}},{\tilde {1}}, {1/2}^+,\pi )}$  
\cr
\noalign{\hrule}
%\noalign {\vskip 12pt}\ & neutrons &  &\  
\cr 1.2670 (35)	 (FREE) & $\ \ $ 0.187 (6)
\cr 0.96 (4) Ref \cite {BW} & $\ \ $ 0.053 (17)
%&\cr
%\noalign{\hrule}
%\noalign {\vskip 12pt}\ & protons &  & \ 
\cr 0.91 (2) Ref \cite {B} & $\ \ $ 0.032 (10) 
\cr 0.891 (FIT) & $\ \ $ 0.024 (6)
\cr
\noalign{\hrule}
\cr EXP Ref \cite {ian2} & $\pm$ 0.024 (1)
\cr 
\noalign {\vskip 12pt}} }
\end {table}
\begin{table}
\caption{Predicted Gamow - Teller strength, $^{39}Ca \rightarrow\  ^{39}K$, for two values of the effective effective axial
coupling constant.}
\vspace{18pt}
\vbox {\tabskip 2em plus 3em minus 1em \halign to \hsize{\hfil  
#\hfill && #\hfil \hfil 
\cr ${{\tilde g}_A} $ & ${\sqrt{B(GT:{\tilde {1}},{\tilde {1}}, {1/2}^+,\nu \rightarrow {\tilde
{1}},{\tilde {1}}, {1/2}^+,\pi} )}$  
\cr
\noalign{\hrule}
%\noalign {\vskip 12pt}\ & neutrons &  &\  
\cr 1.2670 (35)	 (FREE) & $\ \ $ 1.820 (7)
\cr 0.891  & $\ \ $ 1.495 (7)
\cr 
\noalign {\vskip 12pt} }}
\end {table}
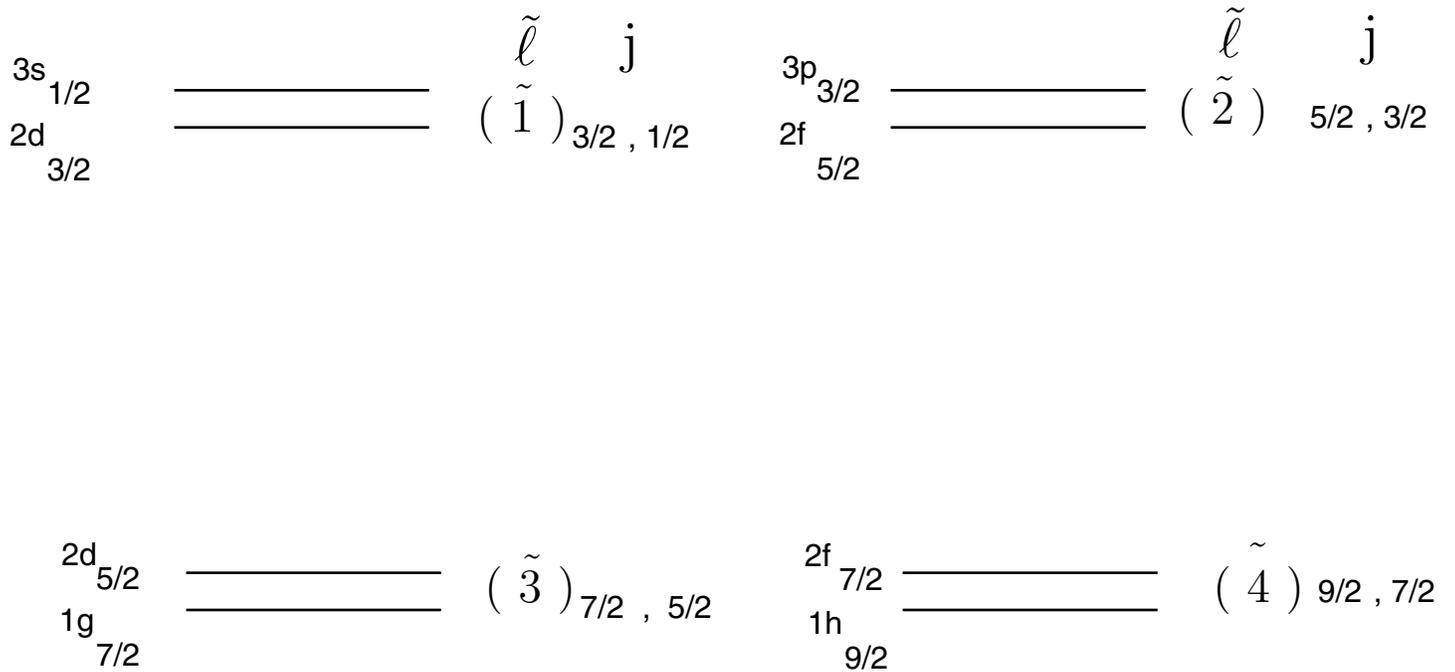
\begin{figure}
\caption{Examples of pseudospin doublets in the $^{208}Pb$ region. $n_r$ is the radial quantum number of the state with $j =
\ell + 1/2 = {\tilde{\ell}} - 1/2$, and is equivalent to $\tilde{n}$, $\tilde{n} = n_r$,
$\ell$ is the orbital angular momentum, $j$ the total angular momentum.}
\label{1}
\end{figure}
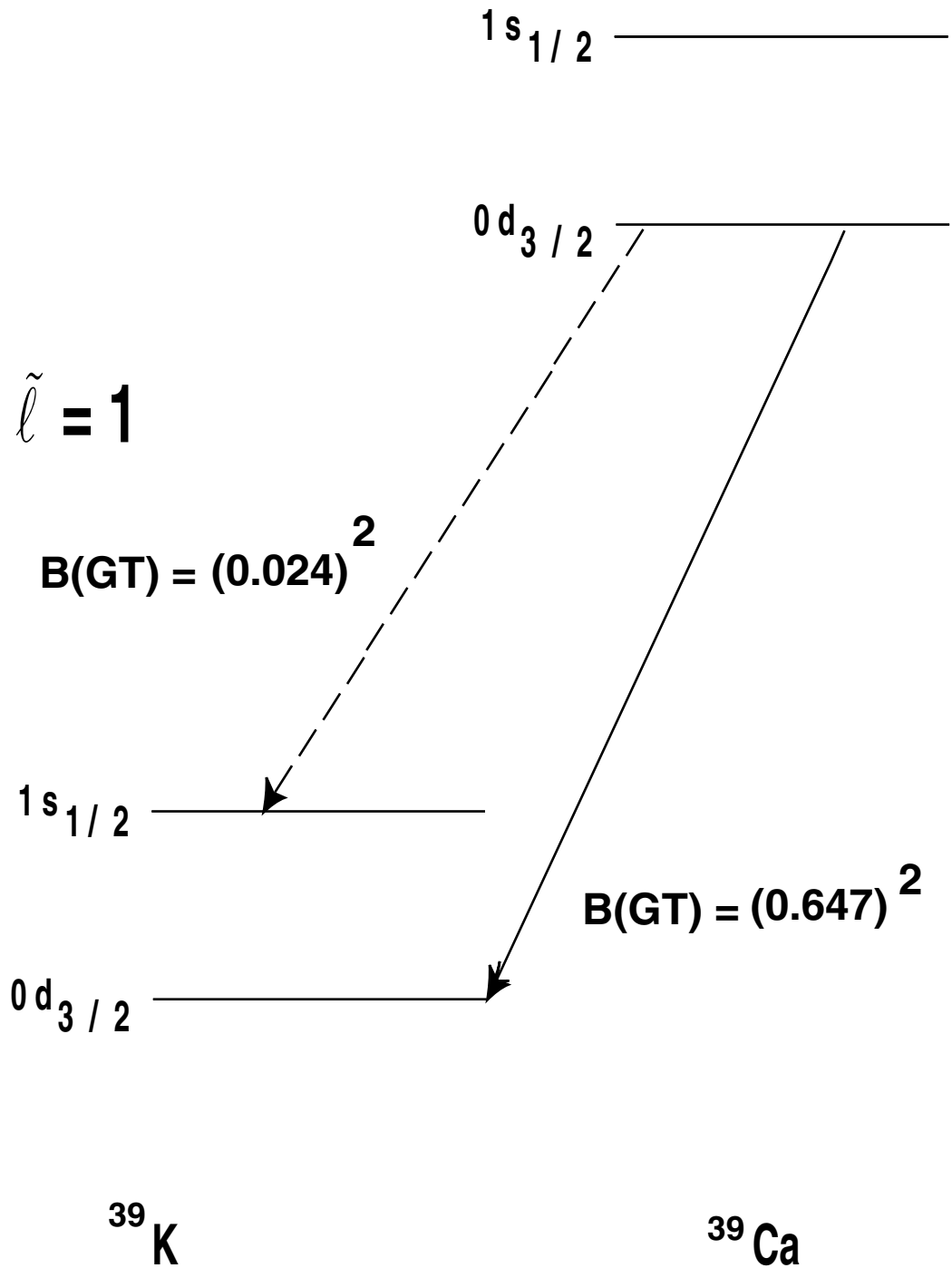
\begin{figure}
\caption{Measured Gamow - Teller transitions between pseudospin doublets for $^{39}Ca$. Dashed line is for ``$\ell$ forbidden''
transition}.
\label{2}
\end{figure}
\pagebreak
\HideDisplacementBoxes
\hskip1.0truein 
\BoxedEPSF{gtpseudospin.eps scaled 800}

\HideDisplacementBoxes
\hskip1.0truein 
\BoxedEPSF{gtca39.eps scaled 800}
\end{document}